# Effects of background solar wind and drag force on the propagation of coronal mass ejection driven shock

Chin-Chun Wu[1], Kan Liou[2], Brian E. Wood[1], and Lynn Hutting[1]

[1] Naval Research Laboratory, Washington, DC 20375, USA
[2] Applied Physics Laboratory, Johns Hopkins University, Laurel, Maryland, USA

## Abstract

Propagation of interplanetary (IP) shocks, particularly those driven by coronal mass ejections (CMEs), is still an outstanding question in heliophysics and space weather forecasting. Here we address effects of the ambient solar wind on the propagation of two similar CME-driven shocks from the Sun to Earth. The two shock events (CME03: April 3, 2010 and CME12: July 12, 2012) have the following properties: Both events (1) were driven by a halo CME (i.e., source location is near the Sun-Earth line), (2) had a CME source in the southern hemisphere, (3) had a similar transit time (~2 days) to Earth, (4) occurred in a non-quiet solar period, and (5) led to a severe geomagnetic storm. The initial (near the Sun) propagation speed, as measured by coronagraph images, was slower (by ~300 km/s) for CME03 than CME12, but it took about the same amount of traveling time for both events to reach Earth. According to the *in-situ* solar wind observations from the *Wind* spacecraft, the CME03-driven shock was associated with a faster solar wind upstream of the shock than the CME12-driven shock. This is also demonstrated in our global MHD simulations. Analysis of our simulation result indicates that the drag force indirectly plays an important role in the shock propagation. The present study suggests that in addition to the initial CME propagation speed near the Sun the shock speed (in the inertial frame) and the ambient solar wind condition, in particular the solar wind speed, are the key to timing the arrival of CME-driven-shock events.

* Work of NRL are supported partially by Office of Naval Research and NASA grant.

# 1. Introduction

Interplanetary (IP) shocks arise from the steepening of magnetosonic waves. Their generation mechanisms and propagation in the solar wind (without collisions) are of great interest in space physics. Within the Earth's orbit, IP shocks are driven predominantly by coronal mass ejections (CMEs). When an IP shock compresses the Earth's magnetosphere and consequently enhances the dayside magnetopause current, it can generate a sudden northward excursion in the geomagnetic field (a.k.a., a storm sudden commencement when it is followed by a geomagnetic storm). Since CMEs, especially those associated with magnetic clouds (MCs) and a large south field in the sheath region, are a major source of geomagnetic storms (e.g., Zhang et al. 2007; Wu et al. 2016), IP shocks are commonly considered a leading indicator of geomagnetic storms. Therefore, understanding how IP shocks (driven by CMEs) propagate in interplanetary space and interact with the solar wind becomes an important subject in space weather research.

The propagation speed of CMEs in the heliosphere can vary significantly, ranging from ~300 to ~1000 km/s or sometimes much more (e.g., Yashiro et al. 2004; Wood et al. 2017). In other words, it takes on average around 18 hours to several days for a CME to reach Earth. Coronagraph images of CMEs from SOHO and/or the STEREO A/B spacecraft allow us to estimate the arrival time of CMEs/shocks at Earth with a significant lead-time window. There are a number of shock-prediction models currently in use, categorized into empirical and physics-based models. In empirical models, the system is simplified such that the shock transit time can be parameterized with some major observables (e.g., Gopalswamy et al., 2000, 2001; Schwenn et al., 2005; Vršnak, 2010). The shock arrival time (SAT, and its driver: ICME or MC) at Earth following the eruption of a CME is a key metric used by those studies. In general, these models provide an averaged prediction error of 10–12 hours (e.g., Owens and Cargill, 2004). A possible prediction error is due to the fact that these empirical models ignore shock dynamical effects. The solar wind is not uniform everywhere and interactions of a CME-driven-shock with the solar wind structures, such as the heliospheric current sheet and the stream interaction regions, could affect the propagation of shocks in the heliosphere. In addition, the solar wind speed from the source region of CMEs/shocks to the observer is typically not observed, adding uncertainties to the modeling, as the drag force is proportional to the square of the velocity difference between the CME and the ambient solar wind $(\Delta v)^2$.

In this study we will study the dynamical effects of the solar wind on Sun-to-Earth shock propagation using numerical magnetohydrodynamic (MHD) simulation. Two Halo CME events will be studied. These two CME events differed significantly in the initial speed but surprisingly took about the same amount of time to propagate to Earth. Description of the two CME events will be given in Section 2. Simulation results and comparisons with shock arrival time at 1 AU in Section 3. Discussion and Conclusions are given in Sections 4 and 5, respectively.

# 2. General Observational Properties of the two CME Events

In order to study the dynamical effect of the solar wind on the CME-driven shock propagation, two fast CME (with an upstream shock) events are selected and studied. The first

CME event occurred on April 3, 2010 (hereafter referred to CME03) and was associated with a B7.4 X-ray solar flare observed at S25E00 at 09:04 UT. It was responsible for the first significant geomagnetic storm (occurred on 5 April 2020) of solar cycle 24 (*e.g.,* Wood *et al.* 2011; Xie et al., 2012; Rouillard et al., 2011). The second CME event took place on July 12, 2012 (hereafter referred to CME12) and was associated with an X1.4 X-ray solar flare erupted at S15W01 at 15:37 UT. In summary, both CMEs (i) were initiated at a similar solar latitude and longitude, (ii) were of the halo type propagating near the Sun-Earth line, (iii) had an upstream shock, presumably driven by the CME, (iv) followed by a large geomagnetic storm after the driver of the shock (ejecta) passed by Earth. The properties of these two CME-shock events are listed in Table 1. The propagation speeds of these two CMEs at different altitudes are estimated using white-light images acquired by the high-cadence, high-resolution *SECCHI* instruments (*Cor1*, *Cor2*, *Hi1*, *Hi2*) (Horward et al. 2008) on board the Solar-Terrestrial Relations Observatory (STEREO) spacecraft. In the following Section we will discuss the white-light image observations of the two CME events in detail.

## 2.1 White-light image observations of CME03 and CME12

Figure 1 shows selected snapshot coronagraph images of CME03 (top panels) and CME12 (bottom panels). CME03 was seen as a compressional wave (presumably a shock) forming ahead of a fast ejecta and emerging into the field of view (FOV) of Cor1a starting at 09:25 UT on 3 April 2010 (~8 o'clock; Figure 1a). At 10:00 UT the CME03's flux rope was clearly visible (see Figure 1b). Figures 1e and 1f show two Cor1a images at 10:25 UT and 16:55 UT on 12 July 2012, respectively. The CME12's flux rope was initiated nearly at the same latitude as CME03 and ejected nearly in the same direction as CME03. Both CME events were also observed by Cor1b (not shown). Figures 1c and 1d show snapshots of the CME03 observed by Cor2a and Cor2b at 11:24 UT, respectively, and Figures 1g and 1h show snapshots of CME12 observed by Cor2a and Cor2b at 17:54 UT, respectively. Both CMEs are similar in terms of their shape and spatial coverage. Note that white-light intensity is normally used for measuring CME's propagation. Running difference method will be adapted if the white-light images are not good enough to identify the CME leading edge (or front boundary) [e.g., Hess and Zhang, 2014; Shi et al. 2015; Mostl et al. 2014]. A GCS model [Thernisien et al. 2011] can be applied to determine the CME initial speeds in a three-dimensional perspective [e.g., Shi et al. 2015; Mostl et al. 2014] when combined with the running difference method. Running difference method is not used in this study.

From a sequence of these coronagraph images, one can estimate the propagation speed of the CMEs within the FOV of each telescope by linearly fitting the height-time data. Such a fitting represents the average propagation speed of the CME within the telescope FOV projected onto the sky plane. Based on this approach, the propagation speed of CME03 ($<V_{CME03}>$) was estimated to be 569, 621, 867, 845 km/s in the FOV of *Cor1a* (Figure 2a), *Cor1b* (Figure 2b), *Cor2a* (Figure 2c), and *Cor2b* (Figure 2d), respectively. Notice that the fitted speed from STEREO-A data is different from STEREO-B data. We will use the average value for the propagation speed. It is shown that CME03 was faster in the FOV of *Cor2* than in the FOV of *Cor1*, increasing from 595 km/s to 855 km/s (or a 44% increase). Thus, it suggests that CME03 was accelerated initially during this time period.

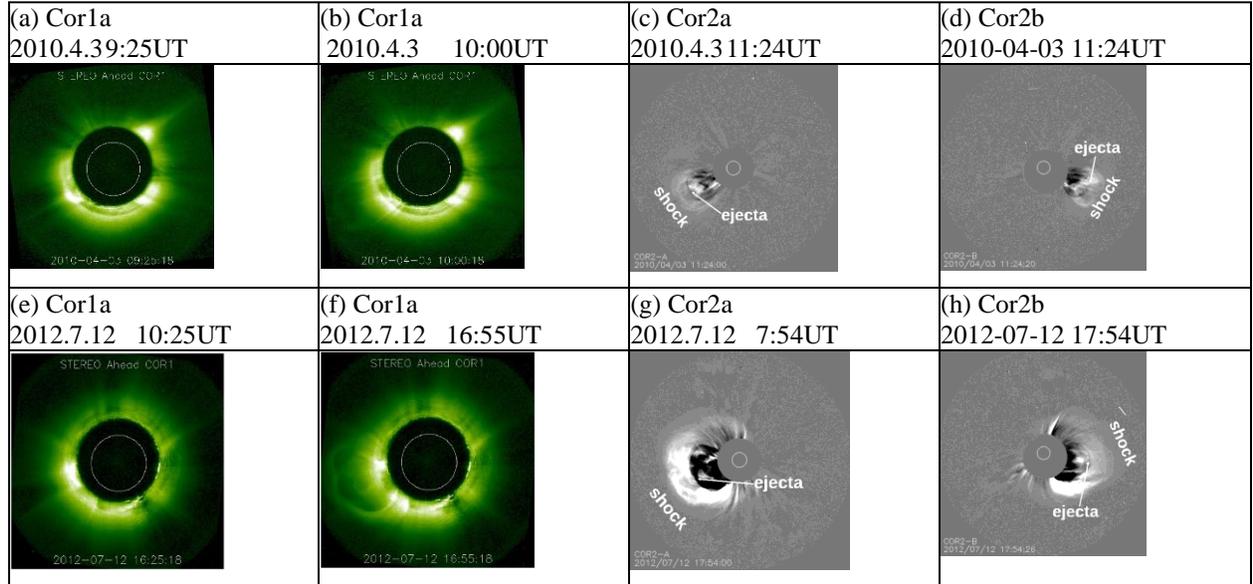

Figure 1. Selected coronagraph images recorded by *STEREO/SECCHI* for (a) CME03 on 03-April 3-4, 2010. At 09:20 UT and for CME12 on July 12-14, 2012 (see text for details). The field of view (FOV) of Cor1 and Cor2 covers 1.5–4 $R_\odot$ and 2–15 $R_\odot$ from the limb, respectively (Howard et al., SSR, 2008).

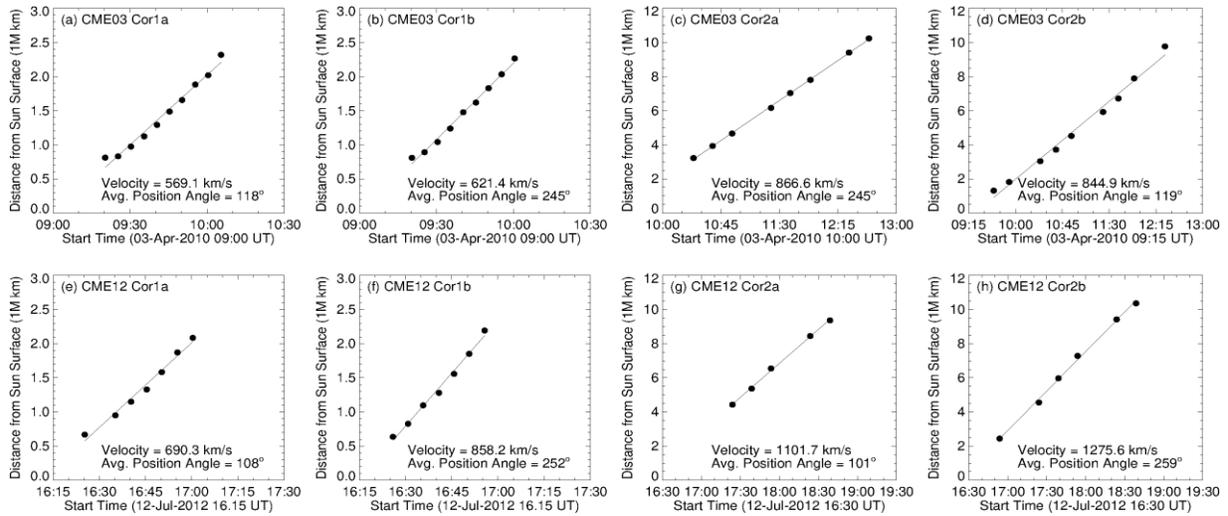

Figure 2. Least-squares regression line representing the average propagation speed of the CME03 (top panels) and CME12 (bottom panels) in the field of view of (a & e) Cor1a, (b & f) Cor1b, (c & g) Cor2a, and (d & h) Cor2b.

The bottom panels of Figure 2 show the estimated propagation speeds of CME12 (<$V_{CME12}$>) within each telescope's FOV. They are 691, 858, 1102, and 1276 km/s in the FOV of *Cor1a* (Figure 2e), *Cor1b* (Figure 2f), *Cor2a* (Figure 2g), and *Cor2b* (Figure 2h), respectively.

The measured speeds used in this study are similar to the previous studies, e.g., 829 km/s for CME03, and 1277 km/s for CME12 [Möstl et al. 2014]; 864 km/s for CME03, and 1224 km/s [Shi et al. 2015]. The difference between this study and the previous studies could be due to points' selection while fitting the measurement for CME's propagation speed. This analysis suggests that the CME12 propagation speed was faster (744 km/s versus 1187 km/s) in the FOV of Cor2 than in Cor1 (or an increase of ~53%). It implies that CME12 was also accelerating during this period (again here we take average values from Cor-a and Cor-b).

We do not apply the same fitting technique to estimate the speeds of CMEs beyond the FOV of Cor2 because only one Hi1 or Hi2 image data exist in either event. It is generally believed that a fast (slow) CME after ejected from the Sun will be decelerated (accelerated) due to the drag force. These two events are clearly in the category of fast CMEs but show acceleration. This is not consistent with the general view. In summary, based on the coronagraph images we determine that the initial speed of CME12 is faster than that of CME03 by 25% in the FOV of Cor1 (1.5–4 Rs) and ~39% in the FOV of Cor2 (2–15 Rs).

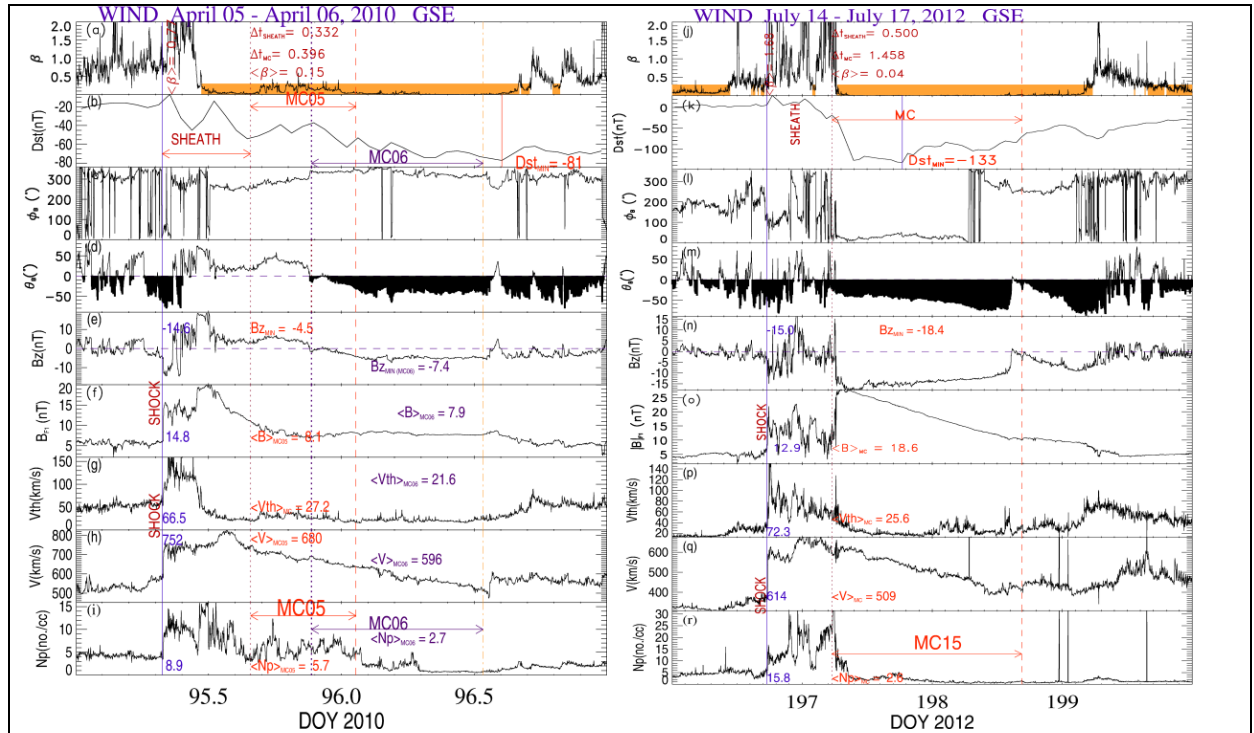

Figure 3. *Wind* observed *in-situ* solar wind parameters and geomagnetic activity index (Dst) during 05-06 April 2010 (left panels) and 13-17 July 2012 (right panels). From Top to Bottom: (**a**) proton plasma beta, (**b**) Dst, , (**c – d**) longitude [$\phi_B$], and latitude [$\theta_B$], in GSE coordinates, (**e**) $B_z$ of the field in GSE, (**f**) magnetic field [$B$] in terms of magnitude, (**g**) thermal velocity [$V_{th}$] or proton temperature [$T$], (**h**) bulk speed [$V$], and (**i**) number density [$N_p$]. The *blue horizontal line* in **panel c** represents the scheme's identification of the extent of this MC candidate (Lepping, Jones, and Burlaga., 1990). Vertical *solid-line*, *dotted-lines*, *dashed-line* represent the IP shock, the front and end boundary of the MC.

## 2.2 *In situ* observations of CME03 and CME12

Now we examine the properties of the two CME events at ~1 AU. Panels on the left of Figure 3 show *in situ* solar wind plasma and magnetic field measurements from the *Wind* spacecraft during April 5–6, 2010 (DOY: 95–96). The *Wind* spacecraft was located at approximately 213 Earth radii ($R_E$) from the Earth during April 3–7, 2010. *Wind* recorded an IP fast-mode shock on April 5, 2010 at ~07:49 UT (refer to Shock05 hereafter) and a shock-like structure at ~11:07UT (referred to Shock-like05 hereafter). After Shock05 encountering *Wind*, a sheath structure followed by a MC at 15:48 UT on April 6 was recorded. Downstream of shock05, the speed was on average 752 km/s. This is usually used for the shock propagation speed. To estimate the travel time for Shock05, we subtract the flare eruption time (April 3, 09:04 UT) from the shock arrival time (April 5, 7:49 UT), this yields 46.8 hrs.

Figure 3, right panels from top down show the *in-situ* solar wind plasma and magnetic field measurements from *Wind* during July 13–17, 2012 (DOY:195–199). *Wind* recorded an IP fast shock at 17:30 UT on July 14, 2012 (referred to Shock14, hereafter). The associated MC (referred to MC15, hereafter) started at 05:28 UT on July 15, 2012 (dotted red line) and ended at 16:31 UT on July 16, 2012 (dashed red line), with a duration of ~35 hours long. This is significantly longer than the ~20 hour duration for typical MCs (*e.g.*, Lepping et al. 2015; Wu & Lepping, 2016). The Shock14 propagation speed was 614 km/s (using the average speed downstream of the shock) and the shock travel time was estimated to be 49.4 hrs. Characteristics of the solar wind parameters in the sheath and driver region for CME03/shock05 and CME12/Shock14 are summarized in Table 1.

Comparing the two CME properties, we found that in general there are some similarities and some differences. For example, Sheath14 is denser and hotter but slower than Sheath05. The magnetic field intensity and its southward component are similar for the two Sheaths, resulting in a similar value of Dst (-40 nT). For the cloud part, the combined MC05 and MC06 (i.e., $MC_{05+06}$) is denser and faster (similar temperature) than CME12. Combing with a larger magnetic field intensity (130%) and it's southward component (200%), CME12 results in a larger storm.

Table 1. Solar wind parameters for the events of CME03-shock05 and CME12-Shock14 and their drivers (MC).

|  | $<N_p>$ | $<V>$ | $<V_{th}>$ | duration | MC type | $Bz_{min}$ | $|B|$ | $Dst_{min}$ |
|---|---|---|---|---|---|---|---|---|
| units | no/cc | km/s | km/s | day |  | nT | nT | nT |
| Sheath05 | 8.9 | 752 | 66.5 | 0.332 |  | -14.6 | 14.8 | -40 |
| MC05 | 5.7 | 680 | 27.2 | 0.340 | N-S | -4.5 | 8.1 | -50 |
| MC06 | 2.7 | 596 | 21.6 | 0.646 | All S | -7.4 | 7.9 | -81 |
| $MC_{05+06}$ | 4.2 | 638 | 24.4 | 0.946 |  | -6.0 | 8.0 |  |
| CME03 | 09:20:18 (*Cor1a*) |  |  | 09:54 (*Cor2a*) |  |  | $V_{CME03-Cor2a}$ | 866.6 km/s |
| Shock05 | 07:49 (*Wind*) |  |  |  |  |  |  |  |
| $\Delta t_{CME03-Shock05}$ |  | 46:28:42 |  |  |  |  |  |  |

| in the group of the fast CME V$_{CME03-Cor2a}$ | 866.6 | | | | | | | |
|---|---|---|---|---|---|---|---|---|
| | | | | | | | | |
| Sheath14 | 15.8 | 614 | 72.3 | 0.50 | | -15.0 | 12.9 | -40 |
| MC15 | 2.6 | 509 | 25.6 | 1.458 | All S | -18.4 | 18.6 | -133 |
| CME12 | 16:25 (*Cor1a*) | | 16:54 (*Cor2a*) | | | V$_{CME12-Cor2a}$ | 1101.7 km/s | |
| Shock14 | 17:00 (*Wind*) | | | | | | | |
| Δt $_{CME12-Shock14}$ | 48:35:00 | | | | | | | |
| V$_{CME12-Cor2a}$ | 1102 | | | | | | | |

**2.3 Discussion of the data analysis**

A fast/slow CME may decelerate/accelerate while the CME was propagating from the Sun to Earth (e.g., Gopalswamy et al. 2000). Both CME03 and CME12 are fast CMEs. It is expected that both CME03/Shock05 and CME12/Shock14 would decelerate right after ejecting from the Sun during outward propagation approaching the solar wind speed. However, the coronagraph images show both CMEs accelerating from the FOV of Cor1 to Cor2. Within the FOV of Cor2 the average CME speed was 855 km/s for CME03 and 1187 km/s for CME12. If both CMEs kept their speed after leaving the FOV of Cor2, it will take ~44.53 hours for CME03/shock and ~32.68 hours for CME12/shock to reach the Wind spacecraft (Wind was located at ~0.9907 AU on April 4, 2010 and 1.0077 AU on July 12, 2012 from the Sun). Note that we have ignored the time for the CMEs propagating from the surface of Sun to the inner FOV of Cor2 at 2 $R_\odot$. Applying the average speed of 595 km/s for CME03 and 744 km/s for CME12, the traveling time is estimated to be 0.32 hr for CME03 and 0.26 hr for CME12. Therefore, the total traveling time from the Sun to *Wind* should be ~44.85 hr for CME03 and ~32.93 hr for CME12. However, this estimate is not in agreement with the observations. The traveling time for CME03/Shock05 and CME12/Shock14 from the Sun (~1 $R_\odot$) to the *Wind* spacecraft is approximately the same, e.g., 46.47 hours and 48.58 hours, respectively. This simplified estimate suggests that after leaving the FOV of Cor2 CME03 must have kept nearly its original speed, whereas CME12 must have undergone significant deceleration. Indeed, a quick calculation indicates that the velocity decrease due to the deceleration is small (a few percent) for CME03 and substantially large (~45%) for CME12. The *Wind* data, which shows 752 km/s and 614 km/s for Shock05 and Shock14, respectively, seems to agree with such an argument.

As mentioned earlier, a general perception is that all fast CMEs will slow down after leaving the Sun. Thus the slow down of CME12 is expected, presumably due to the drag force. However, why CME03 did not slow down noticeably as expected? This question cannot be easily answered because the background solar wind inside of Earth's orbit for the two events was not measured. In situ solar wind measurements at 1 AU may be a good indicator. According to the Wind data, the background solar wind speed upstream of the shock was in the range of 500-600 km/s for Shock05 and 300-350 km/s for Shock14. Since the drag force is proportional to the square of the velocity difference between the CME and the background solar wind, one would expect that the drag force should be smaller for CME03 (ΔV ~550 km/s) than CME12 (ΔV ~650 km/s). Although this is consistent with our previous estimate, one should expect that the large speed difference (thus drag force) on CME12 should also cause the CME03 to be decelerated. Something must have happened that cannot be explained by this simple reasoning. We conclude that solar wind dynamical effects must be very different for the two CME events. In the following Section, we will attempt to use global MHD simulations to help understand the cause and to provide an explanation.

## 3. Global Three-Dimensional MHD Simulation Model (G3DMHD) and Discussion

To study the dynamical effect of the solar wind on the propagation of the two CME events, a fully three-dimensional (3D), time-dependent, magnetohydrodynamic (MHD) simulation code (G3DMHD; Wu et al. 2020a,b), which is an improved version of Han's model (Han 1977; Han, Wu, and Dryer, 1988), was used to propagate solar wind parameters at the inner boundary to 1 AU. The MHD model solves a set of ideal-MHD equations using an extension scheme of the two-step Lax–Wendroff finite-difference methods (Lax and Wendroff, 1960). An ideal MHD fluid is assumed in G3DMHD, which solves the basic conservation laws (mass, momentum, and energy), and the induction equation to take into account the nonlinear interaction between plasma flow and magnetic field.

$$\frac{D\rho}{Dt} + \rho \nabla \cdot V = 0 \quad (1)$$

$$\rho \frac{DV}{Dt} = -\nabla p + \frac{\nabla \times B \times B}{\mu_0} - \rho \frac{GM_s(r)}{r^2} r \quad (2)$$

$$\frac{\partial}{\partial t}\left(\rho g + \frac{1}{2}\rho(V)^2 + \frac{(B)^2}{2\mu_0}\right) + \nabla \cdot \left(V\{\rho e + \frac{1}{2}\rho(V)^2 + p\} + \frac{B \times (V \times B)}{\mu_0}\right) = -V \cdot \rho \frac{GM_s(r)}{r^2} r \quad (3)$$

$$\frac{\partial B}{\partial t} = \nabla \times (V \times B) \quad (4)$$

where $t$, $r$, $\rho$, $V$, $B$, $p$, $e$ are time, radial distance, density, velocity, magnetic field, thermal pressure, and internal energy ($e = p/[(\gamma-1)\rho]$). Additional symbols $\gamma$, $M_\odot$, $G$, $\mu_o$ are the polytropic index, the solar mass, the gravitational constant, and the magnetic permeability in vacuum. $\gamma = 5/3$ is used. The MHD governing equations are cast on uniform, spherical grids. The computational domain for the 3D MHD simulation is a Sun-centered spherical coordinate system $(r, \theta, \phi)$ oriented on the Ecliptic plane. Earth is located at $r = 215\ R_\odot$, $\theta = 90°$, and $\phi = 180°$. The domain covers $2.°5 \leq \theta \leq 177.5°$; $0° \leq \phi \leq 360°$; $18\ R_\odot \leq r \leq 345\ R_\odot$. Open boundary conditions at $\theta = 2.5°$, $\theta = 177.5°$, and $r = 345\ R_\odot$ are used, so there are no reflective disturbances. A constant grid size of $\Delta r = 0.3\ R_\odot$, $\Delta \theta = 5°$, and $\Delta \phi = 5°$ is used, which results in 1100×36×72 grid sets.

### 3.1 Inner Boundary Data Setup

The MHD system is solved with a time-varying inner boundary condition. The photospheric magnetic maps from daily solar photospheric magnetograms (wso.stanford.edu) is used to extrapolate the magnetic field to 2.5 $R_\odot$ using the potential-field source-surface (PFSS) model (*e.g.* Wang and Sheeley, 1992). The location of inner boundary of the G3DMHD model is set at 18 solar radii (R$_\odot$), which is beyond the critical point. The law of conservation of magnetic flux ($r^2B_r$ = constant) is used to derive the magnetic field from 2.5 R$_\odot$ out to 18 R$_\odot$. A formula $V_r = 15_1 + 500 f_s^{-0.4}$ (units in km s$^{-1}$) is used to compute $V_r$ at 18 R$_\odot$, $f_s$ is the expansion factor (Wang and Sheeley, 1990, 1992; Wang et al., 1990). Hereafter, $V_r$ is referred to $V$ for simplicity. Detail information for this methodology can be found in the these studies (Wu et al. 2020a, 2020b). The mass conservation, $\rho V = \rho_o V_o$ = constant, is used to compute the non-uniform, initial solar wind density profile at 18 R$_\odot$ (where $\rho_o$ is assumed to be 2.35 × 10$^{-9}$ kg km$^{-3}$ and $V_o$ is the average value of $V$). We further assume that the total pressure is constant along the stream line (Bernoulli's principle), e.g., $\rho(RT + V^2/2) = \rho_o(RT_o + V_o^2/2)$. This enables us to compute the solar wind fluid temperature at 18 R$_\odot$ where $T_o = 1.5 \times 10^6$ °K is assumed at 18 R$_\odot$. A similar work has been performed previously by McGregor *et al.* (2011). However, they focused on fine-tuning the WSA formula. McGregor *et al.* (2011) adopted a more complicated ($f_s$, $\theta_b$) formula (Equation 2 of McGregor *et al.*) in conjunction with the ENLIL model (Odstrcil, 2005) to specify solar wind speed at the inner boundary set at 0.1 AU along the $\theta$-direction. The background solar wind speed at the inner boundary (18 R$_\odot$) is estimated with an empirical velocity formula, V = 150 + 500 $f_s^{-0.4}$ km/s.

### 3.2 Simulation results

The MHD model described above is used to establish the characteristics of the steady, quiescent solar wind for the CME03 and CME12 event periods. The CMEs are simulated by a pressure pulse injecting at the inner boundary at the time and location of the flare associated with the events (i.e., at 9:04 UT, at a position S23W11 relative to the Sun–Earth axis for CME03). The pressure pulse is formulated as an exponential rise in the wind velocity to a peak value, followed by an exponential decay back to the original value. These peak values are obtained from the FOV of Cor2a to 18 R$_\odot$, i.e., 867 km s$^{-1}$ for CME03 and 1102 km/s for CME12 (the apparent speed in the FOV of C2 is 668 km/s for CME03 and 885 km/s for CME12). The duration of the exponential rise/decay is a free parameter. We adjust the duration to match the arrival time of the simulated CME at Wind with the Wind data. The duration is 5 hours for the CME03 event and 1.75 hours for the CME12 event. In addition, distances between source location and the Wind spacecraft for CME12 event is ~3.6 R$_\odot$ longer than for the CME03-event, because it takes ~0.64 more hours for CME12 to propagate from the Sun to the *Wind* spacecraft.

Figure 4 shows the comparison of MHD simulation result (black traces) with the Wind data (red dots) for the CME03 (*left panels*) and CME12 events (*right panels*). While there are instances where differences exist, the simulated overall solar wind parameters are in reasonably agreement with each other. In addition, the peak values of solar wind parameters downstream of the shock are in good agreement with the Wind data. Notice the perfectly matched shock arrival time for both events. This is done in purpose as this study is trying to understand the propagation of the two CEM events within 1 AU. The simulated fast-mode shock Mach number near the *Wind* spacecraft is 3.4 and 3.6 for the shocks driven by CME03 and CME12, respectively. These values are close to the Wind data (2.6 for CME03 and 3.3 for CME12, see

http://ipshocks.helsink.fi).

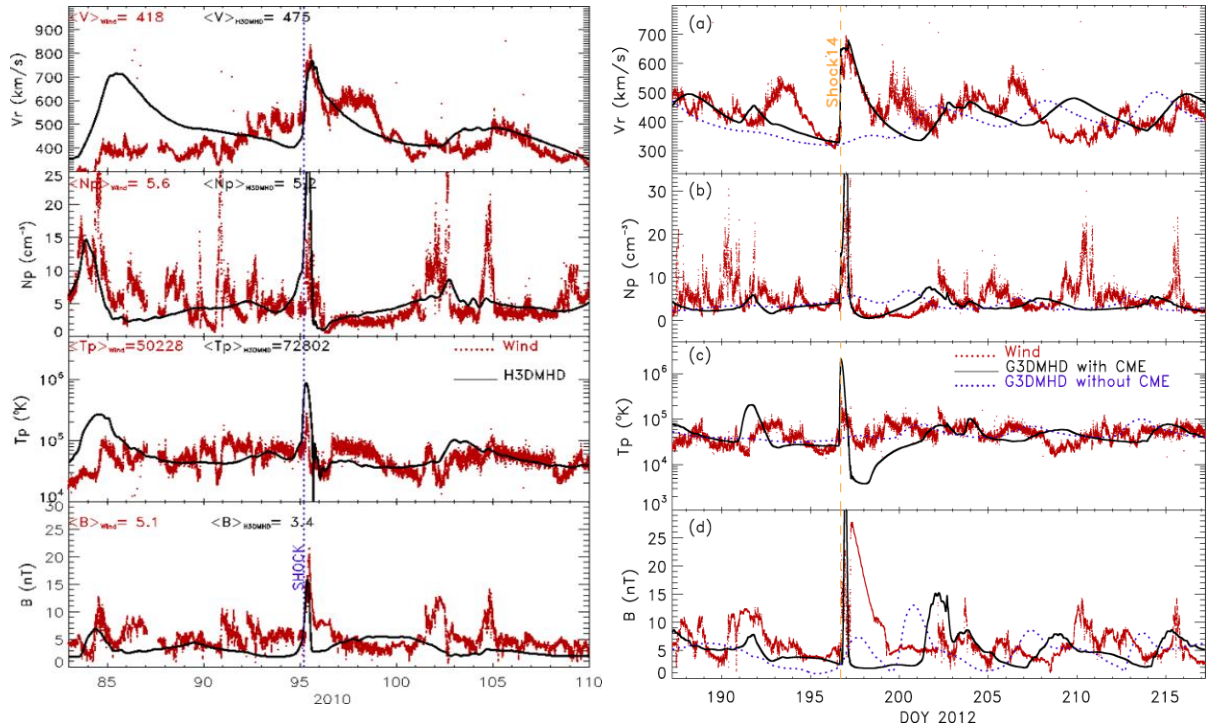

Figure 4. Comparison of the solar wind parameters (from top to bottom panels, *Vr*: radial speed, *Np*: proton density, *Tp*: proton temperature, and *B*: magnetic intensity) between the G3DMHD result (black traces) and Wind *in-situ* (*red-dotted curves*) measurements for the (left panels) CME03 and (right panels) CME12 event periods. The shock arrival time is marked as a vertical dotted line in each panel.

As for the global context, Figure 5 shows four simulated solar wind speed ($V_r$) maps on the surface of angular cone defined by 7.5°S co-latitude or $\theta = 97.5°$ (Sun centered) for the CME03 event and by 7.5°N co-latitude or $\theta = 82.5°$ for the CME12 event. These surfaces contain the Earth's orbit at the event times. Figures 5a and 5e show the solar wind velocity maps ~1 hr after the eruption of CME03 and CME12, respectively. At this time, the background solar wind is mostly undisturbed and is characterized by two (four for CME12) magnetic sectors as indicated by the two (four) heliospheric current sheet determined by $B_\phi = 0$. As intended and expected, the simulated CME was heading toward the Earth. There is an apparent difference in the solar wind speed along the CME trajectory – the background solar wind speed ahead of the CME is much faster for CME03 than for CME12. In addition, CME12 showed a noticeable slowdown (CME color from dark red to light red), whereas CME03 did not.

Effects of background solar wind on the propagation of CME/Shock leading edge is visible which can be seen clearly in Figure 5. Leading edge of the CME/Shocks are deflected and toward the region with higher background solar wind speed. The shock nose (indicated by an arrow) location is visually determined and is marked by black-dashed-circles and pointed by an arrow in each panel of Figure 5. Nose of CME03-Shock05 shifted from 10 o'clock direction to 8:45 o'clock direction. Leading edge of CME-driven-shock are deflected toward the high speed region. In contrast that the nose of CME12-Shock14 is barely shifted.

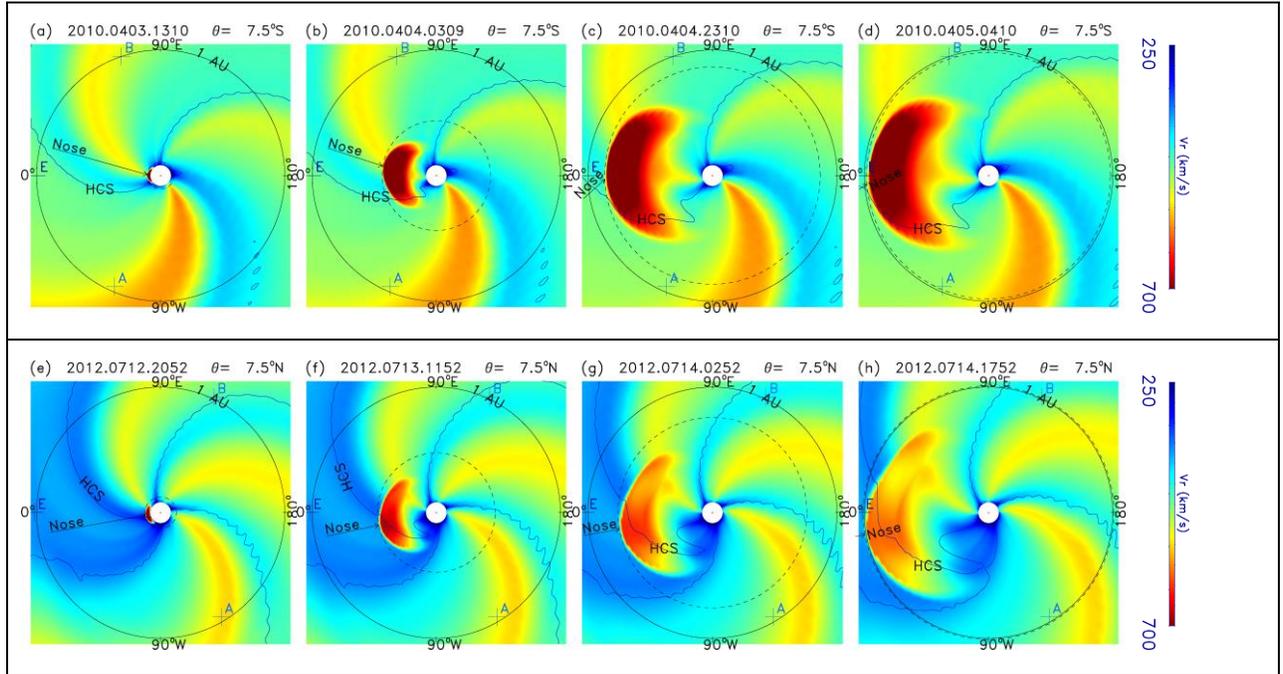

Figure 5. Selected solar wind radial velocity maps on the surface where the Earth is located. The surface is defined by (a–d) $\theta = 97.5°$ for the CME03 and by (e–h) $\theta = 82.5°$ for the CME12 event. The Sun is marked as a red dot in the center. The black circle marks r = 1 AU and the blue contours are the heliospheric current sheet, HCS ($B_\phi = 0$). HCS is marked for the HCS that had been interacting with CME. Nose of the CME-driven-shock are marked as "Nose" with an arrow to the leading edge. Location of spacecraft are marked by a "+" sign, and "E" for Wind, "A" for STEREO-A, and "B" for STEREO-B spacecraft, respectively. Color bar is listed on the right. Black solid-circle indicates the location of 1 AU. Black-dashed-circle indicates the front boundary of the Nose of the CME-driven-shock.

For both CMEs interaction with the HCS occurs along their earthbound transit. However, the interaction seems to be more direct and significant for CME12 than for CME03. This can be easily seen at the nose part of the CME. The CME03 nose did not "touch" the HCS until it hit the Earth. Whereas the CME12 nose interacted with the HCS as early as the time it left the inner boundary. We will present more detailed result and discuss the implication later.

To provide the detailed and quantitative result of our simulation, we plot the solar wind and shock parameters upstream of the shock along the CME trajectory toward the Earth (i.e., along the CME-Earth line) in Figure 6. Panels from top to bottom in Figure 6 show the time profile of simulated magnetic field intensity ($B$), solar wind density ($N_P$), temperature ($T_P$), radial speed ($V_r$), MHD wave-mode speeds ($C_s$: slow mode, $C_A$: Alfvén mode, and $C_f$: fast mode), shock speed in the inertial frame ($V_s$), shock speed in the solar wind frame ($\underline{V_{S@SW}}$), fast-mode Mach number ($M_f$), and radial distance from the Sun's center ($r$) at regions immediately upstream of the shock. For convenience, we will call $r$ as the shock location along the Sun-Earth-line.

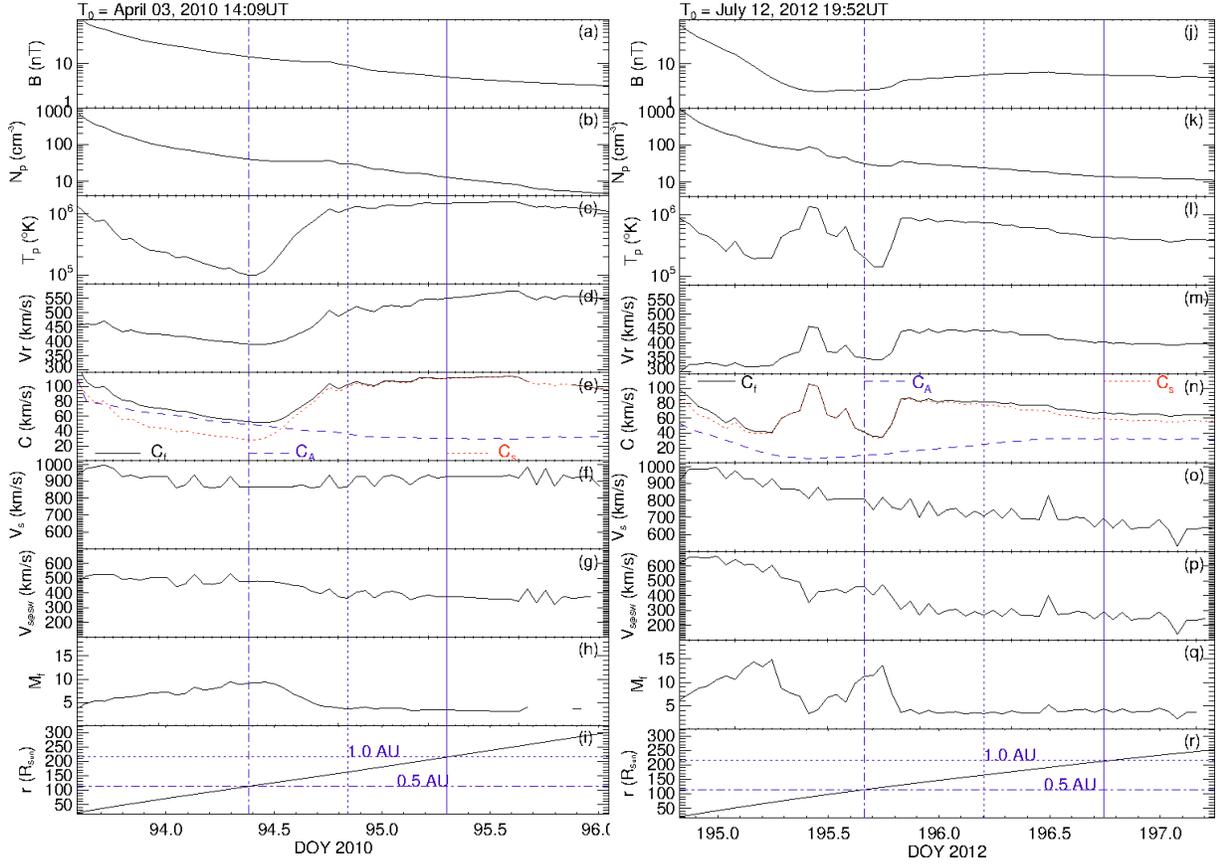

Figure 6. Panels from top to bottom show the time profile of simulated magnetic field intensity ($B$), solar wind density ($N_P$), temperature ($T_P$), radial speed ($V_r$), MHD wave-mode speeds ($C_s$: slow mode, $C_A$: Alfvén mode, and $C_f$: fast mode), shock speed in the inertial frame ($V_s$), shock speed in the solar wind frame ($V_{S@SW}$), fast-mode Mach number ($M_f$), and radial distance from the Sun's center ($r$) at regions immediately upstream of the shock. The left panels (a-i) are for CME03 and the right panels (j-r) are for CME12.

In general, the solar wind properties upstream of the shock is characterized by two different regions separated by a region of increasing all solar wind parameters at $r$ ~120 $R_\odot$ for CME03 and at $r$ ~150 $R_\odot$ for CME12 (i.e., t~14:10UT on April 04 for CME03 and t ~ 23:52UT on July 12 for CME12). According to Figure 5, this is where the shock interacts with the HCS for both events. The solar wind density was nearly the same from 18 $R_\odot$ to 1 AU. However, the magnetic field intensity is smaller for CME12 than for CME03 within ~3/4 AU from the Sun, resulting a smaller Alfvén speed within this region for CME12 than for CME03. The solar wind density profile (Figures 6(b) & 6(k)) was similar from 18 $R_\odot$ to 1 AU for the two events. However, the magnetic field intensity is much smaller for CME12 than for CME03 within ~3/4 AU from the Sun (Figures 6(a) & 6(j)), resulting a smaller Alfvén speed within this region for CME12 than for CME03 (not shown). The solar wind temperature (Figure 6(c) & 6(l)) was larger for CME03 than CME12, except that there is a temperature increase when the shock was in the region between 80 and 130 $R_\odot$, resulting a larger sound speed and fast-mode speed for CME12 than for CME03, except around $r = 90$ $R_\odot$ (Figures 6(e) & 6(n)). For solar wind speed, CME03 is significantly faster within the entire region upstream of the shock nose, except around $r \sim 90$ $R_\odot$ (Figure 6(d) & 6(m)). Nonetheless, both solar wind time-speed profiles are similar. Note that there is a large increase in the solar wind temperature and speed at $r \sim 90$ $R_\odot$ for CME12. This is due to another HCS crossing.

To explore the shock dynamics, we apply the wave-transit method (Wu et al., 1996, 2016) to derive the shock speed ($V_s$). The result is plotted in Figures 6(f) and 6(j), which shows the shock speed in the inertial frame. It is shown that the shock speed was about the same (~1000 km/s) near the inner boundary for both events. After $r \sim 80$ $R_\odot$, the CME12-driven shock started slowing down to ~700 km/s, whereas the CME03-driven shock retained most of its original speed (~900 km/s). In the solar wind frame, the CME12-driven shock was initially faster than the CME03-driven shock (by more than 100 km/s). After the shock reached $r \sim 90$ $R_\odot$, the CME12-driven shock significantly slowed and became slower than the CME03-driven shock (by ~100 km/s at 1 AU). For CME03 the fast-mode shock Mach number was increasing steadily from 4 at $r \sim 20$ $R_\odot$ to ~10 at $r \sim 120$ $R_\odot$, where the Mach number started decreasing quickly to ~3 and remained nearly constant. However, for CME12 the fast-mode shock Mach number did increase from 6 at $r \sim 20$ $R_\odot$ to 15 at r ~75 $R_\odot$, then dropped to ~3 at $r \sim 90$ $R_\odot$, followed by a quick increase to 14 at $r \sim 120$ $R_\odot$. After this radial distance, the shock Mach number dropped quickly to ~3 within ~10 $R_\odot$ and remained constant after that. The Mach number drop at $r \sim 90$ $R_\odot$ is due to the large increase in the proton temperature (thus thermal speed).

## 4. DISCUSSION

There are two parameters that determine the shock arrival time at an observer: The source location and the shock speed. The source location directly affects the shock arrival time because it determines the point on the shock surface that encounters the observer, thus determines the shock travel distance. The source locations for CME03 and CME12 are S25E00 and S13W15, respectively. During April 03–05, 2010 Earth was located at 7° south (S7) from the equator, whereas during July 12-15, 2012 it was located at 7° north (N7) of the equator. Therefore, the nose of CME12/shock14 was farther away from the Sun-Earth line than it was for the CME03-Shock05. The distance between source location and the Wind spacecraft for CME12 is ~3.6 $R_\odot$ longer than for the CME03. Thus, it takes ~38.4 more minutes for CME12 to propagate from the Sun to the *Wind* spacecraft. This is not the major effect on the shock arrival time at 1 AU.

Another parameter that affect the shock arrival time is the shock speed in the inertial frame (or simply the shock speed). According to Figures 6(f) and 6(o), both shocks reached the maximum speed (~1000 km/s) at $r \sim 50$ $R_\odot$. After that the CME03-driven shock slowed down gradually to ~850 km/s then gradually accelerated once it moved across the HCS into the fast wind region and reached ~920 km/s at 1 AU. In contrast, the CME12-driven shock decelerated all the way to 1 AU to ~700 km/s. This is one of the reasons, in addition to the source location, that it takes a shorter time for CME03 to reach the Earth than for CME12 (46 hr versus 48 hr), even though the initial speed of CME12 is faster than CME03 (1102 km/s versus 867 km/s).

What makes the shock speed so different between the two events? The shock speed in the inertial frame is the sum of the shock speed in the solar wind frame (or intrinsic shock speed) and the solar wind speed. Therefore, the shock speed is dictated by the ambient solar wind condition through which the shock propagates and the shock driver (*i.e.*, the CME). The effect of shock driver is negligible after the CME is fully ejected from the source region. For CME03 this time is DOY~93.80 and for CME12 this time is DOY~194.90, both corresponding to when the shock reached ~50 $R_\odot$ and attained its maximum speed and began to slow down (see Figures 6(b) and 6(k)). After these times, the gravitational force is very small and the force acting on the CMEs is dominated by the drag force.

For convenience, we divide the shock propagation path earthward of ~50 $R_\odot$ into slow and fast solar wind regions. The boundary between the two regions is ~150 $R_\odot$ for CME03 and ~130 $R_\odot$ for CME12, which correspond to the fast increase in the solar wind speed after the shock moved across the HCS. In the slow solar wind region, the CME03 intrinsic shock speed decelerated from ~530 km/s to ~430 km/s and the solar wind speed (upstream of the shock) increased from ~440 km/s to ~500 km/s. Therefore the reduction of the shock in the inertial frame is mainly caused by the reduction of the intrinsic shock speed. In the fast solar wind region (i.e., earthward of 150 $R_\odot$), the shock maintained its speed at ~930 km/s until it reached 1 AU and beyond. This is due to the slight increase in the solar wind speed (from 500 km/s to 550 km/s) and the slight decrease in the intrinsic shock speed (from 530 km/s to 370 km/s). Similarly, for CME12 shock speed reduced significantly from ~990 km/s to 750 km/s in the slow solar wind region. This is due mainly to the large reduction of the intrinsic shock speed (from 650 km/s to 320 km/s), although the background solar wind (upstream of the shock) showed an increase from 330 km/s to 450 km/s. In the fast solar wind region the shock speed marginally reduced from 750 km/s to ~700 km/s at 1 AU. This is due to reductions in both the background solar wind (from 450 to 400 km/s) and the intrinsic shock speed (from 320 to 300 km/s).

Therefore, a large slow down of CME12 in the slow solar wind region is the main reason for its longer transit time (than that of CME03) and it is the large reduction of its intrinsic shock speed is the main reason for the slow down. The shock speed is controlled by the shock strength (e.g., the Mach number). In fluid dynamic (as well as MHD), shock waves are a special type of wave drag in supersonic flow. For a compressible flow, the drag force ($F_D$) scales as the square of the Mach number (i.e., $F_D \sim M_f^2$). For CME03, the fast-mode Mach number ($M_f$) between $r$ = 50 $R_\odot$ and $r$ = 150 $R_\odot$ reached a peak value around $M_f$ = 9. On the other hand, for CME12, the fast-mode Mach number between $r$ = 50 $R_\odot$ and $r$ = 120 $R_\odot$ reached peak values of $M_f$ = 15 at $r$ = 70 $R_\odot$ and $M_f$ = 14 at $r$ = 120 $R_\odot$. Although there is a dip in the Mach number ($M_f$ ~4) at $r$ = 90 $R_\odot$, the overall Mach number and thus the drag force is larger for CME12 than for CME03. We believe the wave drag could be the main cause to the large slow down of CME12.

## 5. Conclusions and Remarks

Propagation of CMEs in the heliosphere constitutes an important aspect of collisionless plasma physics. This study focuses on the propagation of two similar halo CME events using data analysis of coronagraph images and global MHD simulations. By comparing the CME/shock speeds and solar wind parameters upstream of the shocks along the Sun-Earth line, it is concluded that the background solar wind parameters play an important role in the CME/shock propagation. Specifically, the present study provides an explanation, in the context of MHD theory, for reasons why the transit time for a CME with a faster initial speed (CME12) is longer than that for a CME with a slower initial speed (CME03). Our simulation result suggests that the solar wind structure can be highly non-uniform along the path of CMEs.

The HCS is usually located between slow and fast wind regions. When a CME/shock moves across the HCS, it will experience the solar wind velocity gradient across the HCS (or CIR to be more precise). Because the CME/shock rides on the solar wind, the shock propagation speed should increase or decrease with the background solar wind. This intuitive view does not always hold. The shock speed (in the initial frame) also play a significant role in the SAT as it can increase or decrease depending on its background solar wind property (e.g., density and temperature). As demonstrated by the simulation result, when a CME and its driven shock propagate through the non-uniform solar wind, the shock strength can change significantly and changes in the shock-induced wave drag can change the shock speed.

The shock background solar wind cannot be measured unless one can move with the shock. The present MHD simulation demonstrated that the background solar wind structure along the CME/shock track can be very complex and no empirical solar wind model can simulate. This is a major impact of the present study as most, if not all, drag-based models employ a smooth solar wind speed profile. Although our MHD simulation of the non-fluxrope CMEs is not perfect, it provides one of the possible solutions for the solar wind and shock speed profiles.

## 6. Acknowledgement

The simulation results (~7 GB of data) of this study can be obtained through making a request to the lead author. All data used in this study are obtained from the public domain. We thank the WIND, STEREO, and LASCO/SOHO PI teams and the National Space Science Data Center at Goddard Space Flight Center, National Aeronautics and Space Administration (NASA) for management and providing solar wind plasma and magnetic field data, and Kyoto University for providing geomagnetic activity index (*Dst*). We also thank Dr. Y. M. Wang (NRL) who provided derived solar magnetic fields at 2.5 $R_\odot$. The work was supported by Office of Naval Research (CCW & BW), NASA grants of 80HRTR19T0062 (KL & CCW), and 80HQTR20T0067 (BW, KL & CCW). The work of K. Liou was also supported by the NASA grant 80NSSC21K0727 to the Johns Hopkins University Applied Physics Laboratory. The authors thank Drs. Christopher Kung and Sam Cable from Engility/DoD High Performance Computing Modernization Office PETTT program for his technical assistance in parallelizing the G3DMHD code.